\documentclass[journal]{IEEEtran}
\IEEEoverridecommandlockouts
\ifCLASSINFOpdf
\else
\fi
\usepackage{cite}
\usepackage{amsmath,amssymb,amsfonts}
\usepackage{graphicx}
\usepackage[ruled,vlined]{algorithm2e}
\usepackage{textcomp}
\usepackage{xcolor}
\usepackage{pgfplots}
\usepackage{bm}
\usepackage{caption}
\usepackage{siunitx}
\usepackage{lipsum}
\usepackage{multirow}
\usepackage{subcaption}
\usepackage{dblfloatfix}
\usepackage{grffile}
\pgfplotsset{compat=newest}
\usetikzlibrary{plotmarks}
\usetikzlibrary{arrows.meta}
\usepackage[T1]{fontenc}
\usepackage[utf8]{inputenc}
\usepackage{pgfplots}
\usepackage{grffile}
\pgfplotsset{compat=newest}
\usetikzlibrary{plotmarks}
\usetikzlibrary{arrows.meta}
\usepgfplotslibrary{patchplots}
\usepackage{amsmath}
\SetKwProg{Init}{init}{}{}
\SetKw{Kwstep}{step}
\usepgfplotslibrary{patchplots}
\usepackage{amsmath}
\usepackage{mathrsfs}
\usepackage{pgfplots}
\usepgfplotslibrary{groupplots}
\usetikzlibrary{arrows}
\usepackage{pgfplots}
\SetKwProg{Init}{init}{}{}
\def\BibTeX{{\rm B\kern-.05em{\sc i\kern-.025em b}\kern-.08em
		T\kern-.1667em\lower.7ex\hbox{E}\kern-.125emX}}

\definecolor{mycolor1}{rgb}{0.00000,0.44700,0.74100}%
\definecolor{mycolor2}{rgb}{0.85000,0.32500,0.09800}%
\definecolor{mycolor3}{rgb}{0.92900,0.69400,0.12500}%
\definecolor{mycolor4}{rgb}{0.49400,0.18400,0.55600}%
\definecolor{mycolor5}{rgb}{0.46600,0.67400,0.18800}%
\definecolor{mycolor6}{rgb}{0.30100,0.74500,0.93300}%
\definecolor{mycolor7}{rgb}{0.63500,0.07800,0.18400}%

\begin{document}
%
\title{Characterization of Dielectric Materials by
	Sparse Signal Processing with Iterative Dictionary Updates}
%
%
%

\author{Udaya~S.K.P.~Miriya~Thanthrige,~\IEEEmembership{Student Member,~IEEE},
        Jan~Barowski,~\IEEEmembership{Member,~IEEE}, Ilona~Rolfes,~\IEEEmembership{Member,~IEEE},
        Daniel~Erni,~\IEEEmembership{Member,~IEEE},
        Thomas~Kaiser,~\IEEEmembership{Senior~Member,~IEEE} and Aydin~Sezgin,~\IEEEmembership{Senior~Member,~IEEE}
\thanks{U.~S.~K.~P.~M.~Thanthrige, and A.~Sezgin are with the Institute of Digital
	Communication Systems, Ruhr University Bochum, 44801 Bochum, Germany
	(e-mail:udaya.miriyathanthrige@rub.de;
	aydin.sezgin@rub.de).}
\thanks{J.~Barowski and I.~Rolfes are with the Institute of Microwave Systems, Ruhr University Bochum, 44801 Bochum, Germany (e-mail:jan.barowski@rub.de; ilona.rolfes@rub.de).}
\thanks{D.~Erni is with the General and Theoretical Electrical Engineering (ATE), Faculty of Engineering, University of Duisburg-Essen, 47048 Duisburg, Germany (e-mail: daniel.erni@uni-due.de).}
\thanks{T.~Kaiser is with the Institute of Digital Signal Processing, University of Duisburg-Essen, 47057 Duisburg, Germany (e-mail: thomas.kaiser@uni-due.de).}
\thanks{This work was funded by the Deutsche Forschungsgemeinschaft (DFG,
German Research Foundation) – Project-ID 287022738 – TRR 196 (S02 and
M04/02 Projects).}
\thanks{Manuscript received \quad \quad \quad ; revised \quad.}}

\IEEEpubid{\begin{minipage}{\textwidth}\ \centering \textbf{2475-1472 \copyright\  2020 IEEE. Personal use is permitted, but republication/redistribution requires IEEE permission. https://www.ieee.org/publications/rights/index.html for more information.}\end{minipage}
 }

\maketitle

\begin{abstract}
Estimating parameters and properties of various materials without causing damage to the material under test (MUT) is important in many applications. Thus, in this letter, we address this by wireless sensing. Here, the accuracy of the estimation depends on the accurate estimation of the properties of the reflected signal from the MUT (e.g., number of reflections, their amplitudes and time delays). For a layered MUT, there are multiple reflections and, due to the limited bandwidth at the receiver, these reflections superimpose each other. Since the number of reflections coming from the MUT is limited, we propose sparse signal processing (SSP) to decompose the reflected signal. In SSP, a so called dictionary is required to obtain a sparse representation of the signal. Here, instead of a fixed dictionary, a dictionary update technique is proposed to improve the estimation of the reflected signal. To validate the proposed method, a vector network analyzer (VNA) based measurement setup is used. It turns out that the estimated dielectric constants are in close agreement with the dielectric constants of the MUTs reported in literature. Further, the proposed approach outperforms the state-of-the-art model-based curve-fitting approach in thickness estimation.
\end{abstract}


%
\IEEEpeerreviewmaketitle

\section{Introduction}
\label{sec:intro}
\IEEEPARstart{C}{haracterization} of materials is important in many areas such as remote sensing, security, and many more. The non-destructive characterization of material is important since it does not cause damage to the material under test (MUT). In general, free-space measurement methods are widely utilized for the non-destructive characterization of material. Unlike other material characterization methods such as the parallel plate capacitor \cite{1561795}, the free-space measurement methods do not require special requirements regarding the size and the shape of the MUT. The free-space measurement techniques are mainly categorized into reflection and transmission modes. However, the transmission mode is not effective in characterizing highly conductive materials or a sample with a large thickness. Therefore, in many applications, the reflection mode is preferred. The free-space measurements can be done in many ways, such as employing a vector network analyzer (VNA) \cite{7461652} or employing a frequency modulated continuous wave (FMCW) radar
\cite{7324408}.\\ \IEEEpubidadjcol  \IEEEpubidadjcol
The main drawback of the reflection mode is the presence of unwanted multiple reflections between the front surface of the MUT and the antenna as well as the reflections within the MUT. Due to limited bandwidth at the receiver, the received signal is broadened in time. Therefore, the received signals of multiple reflections superimpose with each other. Thus, sophisticated signal processing methods are required to resolve these reflections. Accurate estimation of the amplitudes and phases of the reflections of the MUT are required to estimate the MUT's parameters (i.e., the dielectric constant and the thickness). To this end, the reflected signal from the MUT is modeled as a weighted summation of a time-shifted reference signal. Here, the reflected signal from a thin metal plate is used as the reference signal. By using this relationship, an optimization problem is formulated to estimate the amplitudes and phases of the reflections of the MUT.\\ The contributions of this letter are summarized as follows. We propose sparse signal processing to estimate the properties of the reflected signal from the MUT (e.g., number of reflections, their amplitudes and time delays). In SSP, a dictionary is required to obtain a sparse representation of the signal \cite{1710377}. First, we generate a fixed dictionary using time-shifted versions of the reference signal. Second, we propose an iterative dictionary update algorithm to improve the estimation of the properties of the reflected signal from the MUT. Next, the thickness and the dielectric constant of the MUT are estimated using the properties of the reflected signal. For verification purposes, dielectric constants and thicknesses of Teflon, Polyvinylchloride, and Acrylic glass are estimated by the proposed method. Here, the measurements are done by using a VNA based free-space measurement setup. Furthermore, the proposed method is compared with the state-of-the-art method presented in \cite{8823700}.
\begin{figure}[!t]
	\centering
	\includegraphics[ width=0.55\linewidth]{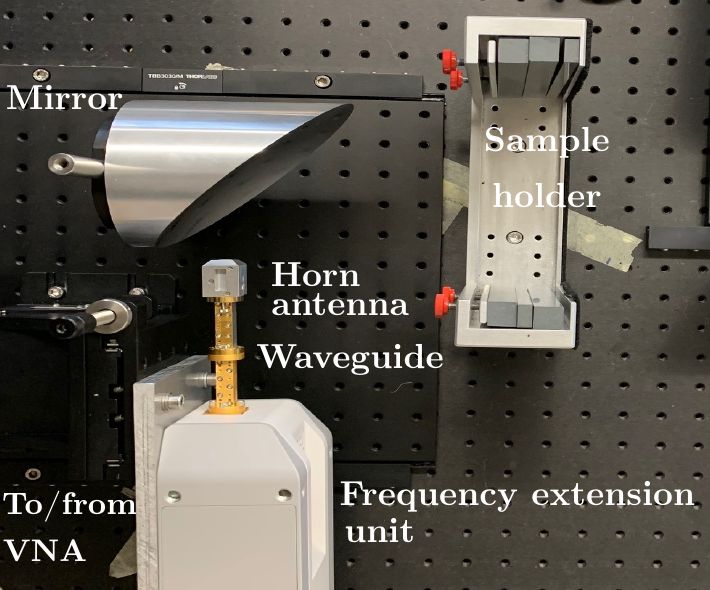}
	\caption{\small Measurement setup with a VNA with frequency extension unit. The MUT is placed in the sample holder (VNA is not shown).}\label{fig:vnasetup}
	\vspace{-0.2cm}
\end{figure}
\section{System Model} \label{smodel}
Consider an active sensing system with a single transmitter and a receiver in mono-static mode. In this work, a VNA is used to generate the transmit signal and measure the reflected signal as shown in Fig. \ref{fig:vnasetup}. Here, the transmitter sends a stepped-frequency continuous wave (SFCW) signal that corresponds to frequency range $f_0$ to $f_0+N\Delta{f}$. Here, starting frequency, the number of frequency steps and the frequency increment are given by $f_0$, $N$ and $\Delta{f}$, respectively. Let the transmit signal at time instant $t$ is given by $x(t)$. Let $K$ be defined as the number of reflections coming from the MUT. Then, the reflected signal from the MUT $y(t)$ is given as
\begin{equation}
\label{equation1}
\begin{aligned}
y(t)=\sum_{k=1}^{{K}}a_{k}~x(t-\tau_k)+ z(t).
\end{aligned}
\end{equation}
\noindent The complex signal strength of the $k-$th reflection and it's time delay are denoted by $a_{k}$ and $\tau_{k}$, respectively.  The $z(t)$ is a zero-mean additive white Gaussian noise with variance $\sigma^2$. Further, $y(t)=h(\tau)*x(t)+ z(t)$. Here, the $*$ and $h(\tau)$ are the convolution operation and the channel impulse response (CIR), respectively. Now, we are interested in the CIR. Because it provides insights about the reflected signal of the MUT (i.e., the number of reflections and their amplitudes and phases). For a linear time-invariant channel, the CIR $\left(h(t)\right)$ is given by
\begin{equation}
\label{cir}
\begin{aligned}
h(t)=\sum_{k=1}^{\text{K}}{a}_{k}~\delta(t-\tau_k).
\end{aligned}
\end{equation}
The $y(t)$ is converted to a fixed intermediate frequency signal $y_{IF}(t)$ by using two harmonics mixers in the VNA. Here, $f_{IF}$ is the intermediate frequency. Now, $y_{IF}(t)$ is sampled by a sampling interval of $T_s$. Thus, the $n-$th sample of the normalized base-band signal is given as
\begin{equation} 
\label{sfcwbbtd}
y_{IF}\left[nT_s\right]=\sum_{k=1}^{\text{K}} {a}_{k} \exp[{-j2\pi(f_{IF}+n\Delta{f})\tau_k}]+ z\left[nT_s\right].
\end{equation}   
Then, the CIR of the MUT is obtained by performing the inverse fast Fourier transform (IFFT) on $y_{IF}$. Next, the CIR is used to estimate the dielectric constant. The frequency ($f$) dependent signal strength of the reflection (${a}_{k}$) depends on many factors. Among them, the free-space attenuation $\left(H_{fs}(f) \right)$ and attenuation in the measurement system $\left(H_{s}(f)\right)$ are dominant. Let the reflection coefficient of the single-layered MUT be $R(f)$. Then, the path-gain of the reflection of the front surface of the MUT $\left({a}_{1}(f)\right)$ is given by
\begin{equation}
\label{equation4}
\begin{aligned}
{a}_{1}(f)=H_{fs}(f)H_{s}(f)R(f)e^{-j2 \pi f \tau_1}=
H(f)R(f)e^{-j2 \pi f \tau_1}.
\end{aligned}
\end{equation}
Suppose that the dielectric constant of the MUT is ${\varepsilon}_{r}(f)$. Then, for normal incidence $R(f)$ is given by
\begin{equation}
\label{equation5}
R(f)=\left(1-\sqrt{{\varepsilon}_{{r}}(f)}\right)~\Big/~\left(1+\sqrt{{\varepsilon}_{{r}}(f)}\right).  \end{equation}
Based on eqs. \eqref{equation4} and \eqref{equation5}, to estimate the $\varepsilon_{{r}}(f)$ of the MUT, $R(f)$ needs to be estimated. However, ${a}_{1}(f)$, $\tau_1$, and, $H(f)$ need to be estimated to obtain the 
$R(f)$. Now, a reference measurement is used to estimate the unknown $H(f)$. For this, a thin metal plate at the MUT's position is used. Alternatively, other objects with known reflection coefficients are suitable as well. The reflection coefficient of the metal plate is $-1$. Thus, the path-gain of the reflection of the front surface of the metal plate (${a}_{1,m}$) is 
\begin{equation}
\label{equation6}
\begin{aligned}
{a}_{1,m}(f) &=H(f)~(-1)~e^{-j2 \pi f \tau_1}. 
\end{aligned}
\end{equation}
Now, by taking the ratio between ${a}_{1,m}(f)$ and ${a}_{1}(f)$ the dielectric constant of the MUT $\left({\varepsilon}_{{r}}(f)\right)$ is estimated.
\begin{equation}
\label{equation7}
\begin{aligned}
R(f)&=-{a}_{1}(f)~\Big/~a_{1,m}(f)={-\tilde{a}}_{1}(f),
\end{aligned}
\end{equation}
\begin{equation}
\label{equation8}
\begin{aligned}
{-\tilde{a}}_{1}(f)&=\left(1-\sqrt{{\varepsilon}_{{r}}(f)}\right)~\Big/~\left(1+\sqrt{{\varepsilon}_{{r}}(f)}\right). 
\end{aligned}
\end{equation}
For simplicity, we write $\tilde{a}_{k}(f)$ as  $\tilde{a}_{k}~\forall~k$ and ${a}_{1,m}(f)$ as ${a}_{1,m}$. Suppose that the metal plate only provides a single reflection. Now, the reflected signal from the metal plate is given as $y_m(t)=a_{1,m}~x(t-\tau_{m,1})+z(t)$. Here, $\tau_{m,1}$ is the time delay of the reflection of the metal plate. Thus, the CIR of the metal plate is given by $h_m(t)=a_{1,m}~\delta(t-\tau_{m,1})$. Therefore, the CIR of the MUT ($h(t)$) can be modeled as a weighted sum of time-shifted CIRs ($h_m(t)$) of the metal plate.
\begin{equation}
\label{equation11}
\begin{aligned}
h(t)=\sum_{k=1}^{{K}}\tilde{a}_{k}~h_m(t-\tilde{\tau}_{m,k}).
\end{aligned}
\end{equation}
Here, $\tilde{\tau}_{m,k}$ is the time delay of the $k-$th reflection of the MUT with respect to the time delay of the metal plate ($\tilde{\tau}_{m,k}=\tau_k-\tau_{m,1}$).
The relative complex signal strength of the $k-$th reflection is denoted by $\tilde{a}_{k}$ and it is given by ${a}_{k}/{a}_{1,m}$. Our main objective is to estimate $\tilde{a}_{1}$ using the CIRs of the MUT and the thin metal plate. Then, the dielectric constant of the MUT is estimated using the eq. \eqref{equation8}. Following, we discuss the estimation of $\tilde{a}_{1}$ using the relationship given in eq. \eqref{equation11}.
\section{Sparse Signal processing}\label{ssp} 
In this section, sparse signal processing (SSP) based estimation of the CIR of the MUT is presented. First, we discuss the fixed dictionary-based approach. Secondly, we introduce an iterative dictionary update algorithm to improve the estimation.
\subsection{Fixed dictionary generation} \label{fixeddl}
Here, we explain the fixed dictionary generation in detail. We discretize the time delay of the reflected signal of the MUT into a uniform grid of size $G$ ($ G \gg K$). This time grid is given as $\bm{\tau}=[\bar{\tau}_{m,1},...,\bar{\tau}_{m,k},...,\bar{\tau}_{m,G}]$ and the size of the grid step is ${\tau_g}=\bar{\tau}_{m,2}-\bar{\tau}_{m,1}$. Now, eq. \eqref{equation11} can be reformulated as
\begin{equation}
\label{equation_ssp_ver}
\begin{aligned}
h(t)=\sum_{k=1}^{{G}}\bar{a}_{k}~h_m(t-\bar{\tau}_{m,k}).
\end{aligned}
\end{equation} 
Let $\bm{\bar{a}}=\left[ \bar{a}_{1},...,\bar{a}_{G}\right]^T$. Since there are only $K$ reflections, the vector $\bm{\bar{a}}~\in \mathbb{C}^G$ is a sparse vector that only has $K$ non-zero elements. The energy of the reflected signal from the MUT decays exponentially in time and is lost in the noise floor. Thus, the CIR can be defined as a finite vector of length $L$. Let the discrete-time CIR of the MUT is given by $\bm{h} \in \mathbb{C}^L$. Let the discrete-time CIR of the metal plate corresponds to time-shift $\bar{\tau}_{m,k}$ is given by $\bm{h}_{m,k} \in \mathbb{C}^L$. Now, the dictionary $\bm{D}_f ~\in \mathbb{C}^{L \times G}$, which contains all the time-shifted CIRs of the metal plate correspond to the $G$ time delays is given as
\begin{equation}
\bm{D}_f=\left[ \bm{h}_{m,1},...,\bm{h}_{m,k},...,\bm{h}_{m,G}\right].
\label{eq:dictionary}
\end{equation}
Note that $k$-th column of the dictionary $\bm{D}_f$ is given by $\bm{h}_{m,k}$ and $\bm{h}_{m,k}$ is the CIR of the metal plate corresponds to the time delay $\bar{\tau}_{m,k}$. Next, we discuss the estimation of the $\bm{\bar{a}}$ using the fixed dictionary $\bm{D}_f$. To this end, the discrete-time version of the eq. \eqref{equation_ssp_ver} is given as  
\begin{equation}
\label{equation13}
\begin{aligned}
\bm{h}= \bm{D}_f \bm{\bar{a}}.
\end{aligned}
\end{equation}
The estimation of $\bm{\bar{a}}$ is formulated as $l^1$-norm minimization problem.
\begin{equation}
\label{equation16}
\begin{aligned}
&\min _{\bm{\bar{a}}} \Vert{\bm{\bar{a}}}\Vert_{1} \\ & \ {\rm s.t.}~ \Vert{\bm{h}-{{\bm{D}_f} \bm{\bar{a}}}\Vert_{2}^{2}~\leq~\epsilon.}
\end{aligned}
\end{equation}
Here, $\epsilon$ is the error tolerance. In this letter, the Orthogonal Matching Pursuit (OMP) \cite{342465} is used to estimate $\bm{\bar{a}}$ in eq. \eqref{equation16}. This method is named as $l^1-$norm minimization based fixed dictionary approach ($FD$). Note that, we cannot guarantee that the time delays of the reflections of the MUT are exactly matched with the grid points of the time grid $\bm{\tau}$. This leads to a grid mismatch. To solve this, we propose an iterative dictionary update algorithm. Next, we discuss this approach.
\subsection{Iterative dictionary update to improve the estimation of CIR of the MUT}
Note that the dictionary $\bm{D}_f$ in Section \ref{fixeddl} is generated based on the time grid $\bm{\tau}=[\bar{\tau}_{m,1},...,\bar{\tau}_{m,k},...,\bar{\tau}_{m,G}]$. Our objective is to update the $\bm{D}_f$ by adjusting these grid points. Let  $\bm{D}=[\bm{d}_{1},...,\bm{d}_{k},...\bm{d}_{G}]$ be the updated dictionary. Here, $\bm{d}_{k}~\in \mathbb{C}^L$ is the $k$-th column of the $\bm{D}$.
The algorithm is initialized by setting $\bm{D}=\bm{D}_f$. Afterward, the iterative dictionary update algorithm alternates between the following steps until we achieve desired convergence criteria $\left(\parallel \bm{h}-{{\bm{D}} \bm{\bar{a}}} \parallel^2_2~\leq~\epsilon \right)$ or the maximum number of iterations is reached:
\begin{itemize}
	\item Step 1: Update $\bm{\bar{a}}$ while keeping the dictionary $\bm{D}$ fixed.
	\item Step 2: Update the dictionary $\bm{D}$ while keeping $\bm{\bar{a}}$ fixed.
\end{itemize}
\begin{figure*}[htbp]
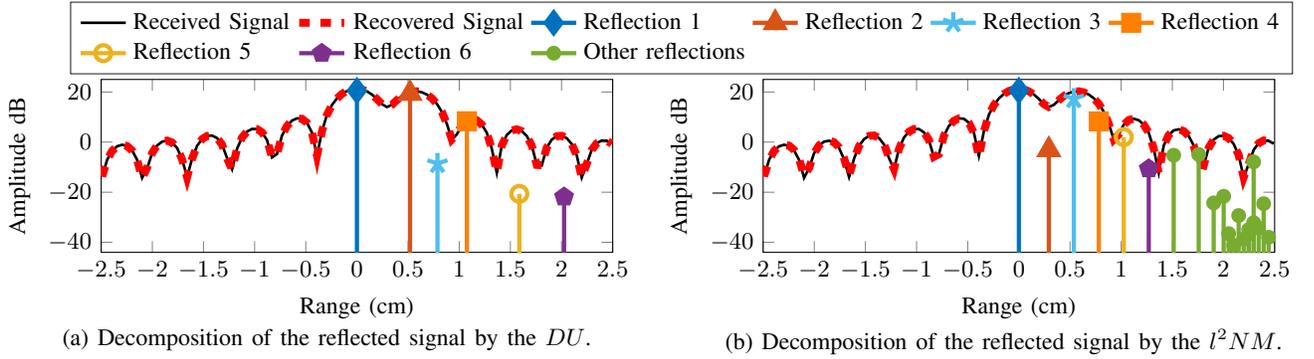

	\centering
	\begin{tikzpicture}
	\pgfplotsset{every tick label/.append style={font=\small}}
	\begin{groupplot}[group style={group size= 2 by 1,horizontal sep = 2cm},]
	\nextgroupplot[width=6.8cm,
	height=2.3cm,
	scale only axis,
	xmin=-0.05,
	xmax=0.05,
	ymin=-44,
	ymax=25,
	legend columns=6,
	xlabel = Range (cm),
	xtick={-0.05, -0.04, -0.03, -0.02, -0.01,  0,  0.01,  0.02,  0.03,   0.04,   0.05},
	xticklabels={$-2.5$, $-2$, $-1.5$, $-1$, $-0.5$,  $0$,  $0.5$,  $1$,  $1.5$,   $2$,   $2.5$},
	ylabel = {Amplitude dB},
	axis background/.style={fill=white},
	title style={font=\bfseries},
	xlabel style={font=\color{white!15!black},font=\small},
	ylabel style={font=\color{white!15!black},font=\small},
	legend style={at={(2.33,1.45)},legend cell align=left, align=left, draw=white!15!black,font=\small,row sep=-0.1pt},scaled ticks=false, tick label style={/pgf/number format/fixed},]
	\input{figures/PMMA_03300L1.tex}
	\nextgroupplot[width=6.8cm,
	height=2.3cm,
	scale only axis,
	xmin=-0.05,
	xmax=0.05,
	ymin=-44,
	ymax=25,
	legend columns=1,
	xlabel = Range (cm),
	xtick={-0.05, -0.04, -0.03, -0.02, -0.01,  0,  0.01,  0.02,  0.03,   0.04,   0.05},
	xticklabels={$-2.5$, $-2$, $-1.5$, $-1$, $-0.5$,  $0$,  $0.5$,  $1$,  $1.5$,   $2$,   $2.5$},
	ylabel = {Amplitude dB},
	axis background/.style={fill=white},
	title style={font=\bfseries},
	xlabel style={font=\color{white!15!black},font=\small},
	ylabel style={font=\color{white!15!black},font=\small},
	scaled ticks=false, tick label style={/pgf/number format/fixed},]
	\input{figures/PMMA_03300L2.tex}
	
	\end{groupplot}
	\node[text width=10cm,align=center,anchor=north] at (3,-0.9) {{\small { (a) Decomposition of the reflected signal by the $DU$}.
			\label{fig:pmma03300L1}}};
	\node[text width=12cm,align=center,anchor=north] at (12,-0.9) {{\small {(b) Decomposition of the reflected signal by the $l^2NM$.}\label{fig:pmma03300L2}}};
	
	\end{tikzpicture}
	\caption{\small {Decomposition of the reflected signal of the PMMA sample with $3.3~$\SI{}{\milli\metre} thickness. Here, the first reflection shown at $0~\SI{}{\centi \metre} $}.}
	\label{fig:pmma03300}
\end{figure*}
In the first step, the OMP algorithm is used to estimate $\bm{\bar{a}}.$ Let $s_0$ be the number of non-zero elements in $\bm{\bar{a}}$. After the first step, the indices of the non-zero elements in $\bm{\bar{a}}$ are known. This index set $\Omega$ is given by $\Omega=\{k~|~\bar{a}_{k}~\neq~0\}$. Note that the OMP algorithm utilizes the correlation between the input $\bm{h}$ and the columns of $\bm{D}$ to obtain the $\Omega$. Motivated by this, in the second step, the correlation between the input $\bm{h}$ and the columns of $\bm{D}$ is used to update the dictionary. It is worth noticing that, only the columns of the dictionary which correspond to the non-zero elements in $\bm{\bar{a}}$ are updated. The iterative dictionary update process is given in  algorithm \ref{algorithm1}. We name this method as $l^1-$norm minimization based iterative dictionary update approach ($DU$).\\
\begin{algorithm}[tb]
	\SetKwInput{KwInput}{Input}                
	\SetKwInput{KwOutput}{Output}              
	\DontPrintSemicolon
	\DontPrintSemicolon
	\KwInput{$\bm{D}_f=[\bm{d}_{1},...,\bm{d}_{q},...\bm{d}_{G}]~\in \mathbb{C}^{L \times G}-$Dictionary\; 
		\hspace{1.0cm} $\bm{\tau}=[\bar{\tau}_{m,1},...,\bar{\tau}_{m,k},...,\bar{\tau}_{m,G}]-$Time grid\;
		\hspace{1.0cm} $\bm{h}~\in \mathbb{C}^{L}-$CIR of the MUT \; 
		\hspace{1.0cm} $s_0-$Number of non-zero components of $\bm{{\bar{a}}}$\;
		\hspace{1.0cm} $\epsilon-$Error tolerance \; 
		
		}
	\bf{Initialize:}\normalfont
	\hspace{0.2cm} $count \gets 1$\;
	\hspace{1.6cm} $\bm{D} \gets \bm{D}_f$\;
	\hspace{1.6cm} $\bm{r} \gets \bm{h}$\;
	\hspace{1.6cm} $er(count) \gets ||\bm{h}||_2^2$\;
	\hspace{1.6cm} ${\tau}_w \gets (\bar{\tau}_{m,2}-\bar{\tau}_{m,1})/2$\;
	\While{$er(count)$ $>$ $\epsilon~ \text{or}~count < P+1$}
	{
		$count \leftarrow count+1$\;
		\textbf{Step 1:} Fixed $\bm{D}$ and update $\bm{\bar{a}}$ \;
		$\bm{\bar{a}}=\left[ \bar{a}_{1},...,\bar{a}_{k},...,\bar{a}_{G}\right]^T \gets \text{OMP}(\bm{D},\bm{h},s_0)$\;
		$\Omega \gets \{k~|~\bar{a}_{k}~\neq~0\}$\;
		$er (count) \gets ||\bm{h}-\bm{D}\bm{{\bar{a}}}||_2^2$\;
		\If {$er(count)$ $\leq$ $\epsilon$} {break}		 
		\textbf{Step 2:} Fixed $\bm{\bar{a}}$ and update $\bm{D}$\;
		\For{$i_1\leftarrow 1$  \KwTo $s_0$ \Kwstep $1$ } 
		{ 
			$j \gets \Omega(i_1)$\;
			\text{Generate a new time grid $\bm{\tau}_{md}$, around $\bar{\tau}_{m,j}$}\; 
			$\bm{\tau}_{md}\gets[\bar{\tau}_{m,j}-\tau_w/2:\tau_{mg}:\bar{\tau}_{m,j}+\tau_w/2]$\;
			\text{Generate a new mini-dictionary $\bm{D}_{md} ~\in \mathbb{C}^{L \times C}$ using} \text{the time grid  $~\bm{\tau}_{md}$ as described in Section \ref{fixeddl}}
			$\bm{D}_{md} \gets[\bm{d}_{md,1},...,\bm{d}_{md,q},...\bm{d}_{md,C}]$\;
			{Find the largest correlation column of $\bm{D}_{md}$ \;
				$I \leftarrow $ $\underset{i}{\mathrm{argmax}}|(\bm{d}_{md,i})^T\bm{r}|, ~ i \in \{1,...,C\}.$\;
				\vspace{0.15cm}
				\If{$|(\bm{d}_{j})^T\bm{r}|~<~|(\bm{d}_{md,I})^T\bm{r}|$}
				{$\bm{d}_{j} \gets \bm{d}_{md,I}$ \quad \ \ \ \tcp{\small update the $\bm{D}$}		
					$\bar{\tau}_{m,j} \gets \bm{\tau}_{md}(I)~$ \tcp{\small update the $\bm{\tau}$}
				}
				
			}
			$\bm{r} \gets \bm{r}-\bm{d}_{j}\bar{a}_{j}$ \quad   \tcp{\small update the residual}
		}
	}
	\KwOutput{$\bm{D}$, $\bm{\tau}$, $\bm{{\bar{a}}}$}
	\caption{The iterative dictionary updates.}
	\label{algorithm1}
\end{algorithm}
To have a fair comparison with the $FD$ and $DU$, we estimate $\bm{\bar{a}}$ using the $l^2$-norm minimization as given below. This method is named as $l^2-$norm minimization approach ($l^2NM$).
\begin{equation}
\label{equationL2}
\begin{aligned}
&\min _{\bm{\bar{a}}}\Vert{\bm{\bar{a}}}\Vert_{2}^{2} \\ & \ {\rm s.t.}~\Vert{\bm{h}-{{\bm{D}_f} \bm{\bar{a}}}\Vert_{2}^{2} \leq \epsilon.}
\end{aligned}
\end{equation}
\section{Measurement Setup and Results} \label{msetup}
For the verification of the proposed method, various materials with different thicknesses as listed in Table \ref{table1} were tested. Actual thicknesses and measured frequency ranges of the MUT's are listed in the first column of Table \ref{table1}. The measurement setup based on the VNA shown in Fig. \ref{fig:vnasetup} was used as the experimental setup. The MUT was placed in the sample holder and the horn antenna radiated the EM waves to the MUT as shown in Fig. \ref{fig:vnasetup}. After the reflected signal of the MUT was measured. Next, the thin metal plate was placed in the sample holder to measure the reference measurement. Here, the IFFT was used to obtain the CIRs of the MUT and the metal plate. Next, SSP based approaches given in Section \ref{ssp} were used to estimate $\bm{\bar{a}}$ of the MUT. Afterward, the relative complex signal strength of the reflection of the front surface of the MUT was used to estimate the dielectric constant. In this work, $s_0$, the number of nonzero elements in $\bm{\bar{a}}$ was changed from $2$ to $8$ with the step-size of $1$ and the $s_0$ value which provides the lowest error ($\left(\parallel \bm{h}-{{\bm{D}} \bm{\bar{a}}} \parallel^2_2\right)$) was selected. The $\epsilon$ was set as $10^{-2}$.
\subsection{Decomposition of the received signal using the $l^1$ and the $l^2$-norm minimization}  
First, we analyze the decomposition of the reflected signal from the MUT by the $l^1$ and the $l^2-$norm minimization approaches. Fig. \ref{fig:pmma03300} (a) and \ref{fig:pmma03300} (b) show the decomposition of the reflected signal of the PMMA sample with $3.3~$\SI{}{\milli\metre} thickness by using the proposed $DU$ and $l^2NM$ approaches, respectively. The amplitudes and the time delays of the estimated reflections by these methods are indicated as vertical lines in Fig. \ref{fig:pmma03300}. The first and the second dominant reflections which are shown in Fig. \ref{fig:pmma03300} correspond to the front and the backside reflections of the MUT. The remaining reflections correspond to the multiple reflections between the antenna and the front surface of the MUT and also between the front and
the back surfaces of the MUT.\\
It can be seen that both $DU$ and $l^2NM$ approaches estimate the reflected signal from the MUT accurately. However, the $l^2NM$ estimates more reflections when compared to the $DU$. Note that, the time delays of the internal reflections of the MUT should be integer multipliers of the time-delay between the first and the second dominant reflections. Most of the time delays of the recovered reflections by the $DU$ are integer multipliers of the time-delay between the first and the second reflection. (e.g., fourth, fifth and sixth reflections shown in Fig. \ref{fig:pmma03300} (a)). However, in the $l^2NM$, this is not the case. Thus, the proposed $DU$ is able to recover internal reflections inside the MUT better than the $l^2NM$. Further, the $l^2NM$ does not accurately estimate the actual number of reflections. Hence, the estimation accuracy of the first and the second dominant reflections by the $l^2NM$ decreases. Thus, it can be concluded that the $l^1-$norm minimization performs much better compared to the $l^2-$norm minimization.
\begin{table}[!t]
	\centering
	\caption{Estimated dielectric constants and thicknesses by the proposed approach ($DU$), the $FD$, the $l^2-$norm minimization  ($l^2NM$) and the curve-fitting approach ($CF$).}
	\centering
	\renewcommand{\arraystretch}{1.3}
	\begin{tabular}{|c|l|c|l|l|l|}
		\hline
		\multirow{2}{*}{MUT}                                                                    & \multirow{2}{*}{Method} & \multicolumn{2}{c|}{\begin{tabular}[c]{@{}l@{}}Estimated \\ Dielectric\\ Constant\end{tabular}} & \multicolumn{2}{c|}{Thickness}                                         \\ \cline{3-6} 
		&                         & $\acute{\varepsilon}_{r}(f)$                                    & \begin{tabular}[c]{@{}c@{}}Loss \\ factor  \\ $\left(\text{tan}~\delta\right)$  \end{tabular}                    & \begin{tabular}[c]{@{}c@{}}Estimated \\ ($\SI{}{\milli\metre}$)\end{tabular} & \begin{tabular}[l]{@{}l@{}}Esti-\\mation\\ error\end{tabular} \\ \hline
		\multirow{4}{*}{\begin{tabular}[l]{@{}l@{}}Acrylic glass \\ (PMMA)\\ $3.30~\SI{}{\milli\metre}$ \\ (75-110 GHz)\end{tabular}} &$FD$&$2.5920$ &$0.0048$&$3.1367$&$4.95$   \\ \cline{2-6}
		&    $DU$ & $2.5990$ &$0.0044$  &  $3.2099$ &  $\bf{2.73}$  \\ \cline{2-6}
		&    $l^2NM$ & $2.6095$ &$0.0331$  &  $3.0951$    &$6.21$  \\ \cline{2-6}
		&  $CF$  &$2.6132$ & $0.0262$&$3.1780$ & $3.70$  
		\\ \cline{2-6}  \hline
		\multirow{4}{*}{\begin{tabular}[l]{@{}l@{}}Acrylic glass \\ (PMMA)\\ $3.81~\SI{}{\milli\metre}$  \\ (220-330 GHz)\end{tabular}}&$\small FD$&$2.5869$&$0.0952$&$4.0399$  &  $6.03$   \\ \cline{2-6} 
		&     $DU$ & $2.5718$ &$0.0968$  &  $3.9115$ &   $\bf{2.66}$  \\ \cline{2-6} 
		&    $l^2NM$ &  $2.3327$ & $0.1115$ &  $3.9262$ &   $3.05$  \\ \cline{2-6}
		&   $CF$  &$2.5672$ & $0.0352$&$3.9160$&  $2.78$    \\ \cline{2-6}   \hline	
		
		\multirow{4}{*}{\begin{tabular}[l]{@{}l@{}}PVC \\ $15.76~\SI{}{\milli\metre}$\\ (75-110 GHz)\end{tabular}}&$FD$&$2.7602$&  $0.1797$&$15.1960$  &  $3.58$   \\ \cline{2-6}
		&    $DU$ & $2.7240$ &$0.1784$  &  $15.2955$ &   $2.95$  \\ \cline{2-6}	
		&  $l^2NM$ &  $2.5989$& $0.1708$ & $15.4910$  &    $\bf{1.71}$ \\ \cline{2-6}
		&   $CF$  &$2.8895$&  $0.0299$&$14.8620$&    $5.70$   \\ \cline{2-6}	  \hline
		
		\multirow{4}{*}{\begin{tabular}[l]{@{}l@{}}Teflon (PTFE) \\ $20.30~\SI{}{\milli\metre}$ \\ (75-110 GHz)\end{tabular}} &$FD$&$2.0006$& $0.0408$& $20.4300$&  $0.64$   \\ \cline{2-6}
		&    $DU$ & $2.0015$ &$0.0406$  &  $20.4217$ &   $\bf{0.60}$  \\ \cline{2-6} 
		&    $l^2NM$ & $1.8023$ & $0.3768$ &  $22.1100$ &    $8.91$ \\ \cline{2-6}	
		&  $CF$ & $2.0582$& $0.0030$& $20.0940$&  $1.01$    \\ \cline{2-6}   \hline
	\end{tabular}
	\label{table1}
\end{table}
\begin{table}[!t]
		\centering
	\renewcommand{\arraystretch}{1.20}
	\caption{Comparison of the real part of estimated dielectric constants by the proposed approach ($DU$) with the literature.}
	\begin{tabular}{|l|l|l|}
		\hline
		{MUT} &  by the $DU$ & From literature                                                                                                                                                                                           
		\\ \hline
		PMMA                 & 2.599, 2.5718
		& \begin{tabular}[c]{@{}l@{}} 2.58-2.60 \cite{8617322}, 2.58-2.61 \cite{ozturk2017development} \end{tabular}                                          \\ \hline
		PVC                  & 2.7240
		& \begin{tabular}[c]{@{}l@{}}2.738\cite{596579}, 			2.83-2.89 \cite{6983634}\end{tabular}                   \\ \hline
		Teflon (PTFE)        & 2.0015                                                                                                 & \begin{tabular}[c]{@{}l@{}} 2.03 \cite{8617322}, 2.02–2.04 \cite{ozturk2017development} \end{tabular} \\ \hline
	\end{tabular}
		\vspace{-4mm}%
	\label{table2}
\end{table}
\subsection{MUT's Dielectric constants and thicknesses estimation }
Next, we are going to discuss the dielectric constants and thicknesses estimation of the MUTs. For comparison, the dielectric constants and thicknesses of the MUTs were estimated by the model-based curve-fitting approach ($CF$) presented in \cite{8823700} and the $l^2-$norm minimization ($l^2NM$). Here, the loss factor $\left(\text{tan}~\delta\right)$ of the MUT is calculated by ${\varepsilon}_{{r}}(f)={\acute{\varepsilon}}_{{r}}(f)(1-\text{tan}~\delta)$. 
Table \ref{table1} shows the comparison of the dielectric constants and thicknesses that are estimated by the proposed $DU$ and also by the other approaches ($FD$, $l^2NM$ and $CF$). Based on the results in Table \ref{table1}, it can be seen that the estimated dielectric constants by the $l^1-$norm minimization ($DU$ and $FD$) and the $l^2-$norm minimization ($l^2NM$) show close agreement for the thin MUTs.\\
As an example, consider the PMMA sample with $3.3~$\SI{}{\milli\metre} thickness. For this sample, the estimated dielectric constants by $DU$, $FD$ and $l^2NM$ approaches show close agreement with the dielectric constants which are reported in the literature. However, for thick samples, the estimated dielectric constants by the $l^2-$ norm minimization approach show higher deviations compared to the values reported in the literature. As an example consider the PVC sample with $15.76~$\SI{}{\milli\metre} thickness. In this case, the estimated dielectric constants by $DU$ and $FD$ are much closer to the dielectric constants which are reported in the literature when compared to $l^2NM$.\\
This is due to the fact that, as sample thickness increases EM waves travel more distance inside the sample in internal reflections. Thus, for a thin sample, losses inside the sample are smaller compared to a thick sample. Therefore, for a thin sample, we could expect more internal reflections than for a thick sample. Thus, the reflected signal of a thin sample is less sparse compared to a thick sample. Due to this reason, the $l^2-$norm minimization provides less accurate results compared to the $l^1-$norm minimization based approaches ($DU$ and $FD$) for thick samples. Note that the $l^1-$norm minimization based approaches ($DU$ and $FD$) show close agreement with the dielectric constants which are estimated by the state-of-the-art $CF$ approach and the dielectric constants given in the literature (as shown in Table \ref{table2}) for all samples.\\
Next, we are going to discuss the thickness estimation of the MUTs. As given in Table \ref{table1}, the proposed $DU$ approach provides the lowest thickness estimation error for the majority of samples (three out of four samples). For the other sample, the $DU$ provides the second-best thickness estimation with $1.2\%$ difference from the best. Moreover, by comparing the thickness estimation error between $DU$ and $FD$, it can be observed that the iterative dictionary updates improve the thickness estimation. Here, the improvements are $2.22\%$, $3.37\%$, $0.63\%$ and $0.04\%$ for the MUTs given in Table \ref{table1}, respectively. Although the thickness estimation improvements from the $FD$ to the $DU$ are small in numbers, the $DU$ is the best thickness estimator for three out of four MUTs. Further, the $FD$ is not the best thickness estimator for any MUT given in Table \ref{table1}. Therefore, it is worth to consider the iterative dictionary updates rather than a fixed dictionary.\\ Finally, we discuss the computational complexity of the $DU$ and the $FD$. For a single iteration of the $DU$, the computational complexities of steps 1 and 2 of the algorithm \ref{algorithm1} are $ O(Ls_0^2+LGs_0)\approx O(LGs_0)$ and $O(CLs_0)$, respectively. Suppose that the algorithm \ref{algorithm1} converges after $P$ iterations. Therefore, the computational complexity of the algorithm \ref{algorithm1} is given by $O\left(PLs_0(C+G)\right)$. Since $G \gg C$, $O\left(PLs_0(C+G)\right) \approx O\left(PLGs_0\right)$. Next, we discuss the computational complexity of the $FD$. The $FD$ utilizes OMP to obtain $\bm{\bar{a}}$. Thus, the computational complexity of the $FD$ is given by $O(LGs_0)$. Further, based on the simulations, we observed that the $DU$ (algorithm \ref{algorithm1}) converges after a few iterations (such as $P < 10$). Thus, typically the computational complexity of the $DU$ is less than ten times of the computational complexity of the $FD$. However, the $DU$ has higher computational complexity, yet it is worth to consider the $DU$ since it improves the thickness estimations of the MUTs.
\section{Conclusion}
In this letter, we investigated the material characterization in free-space reflection mode. Here, a VNA based measurement setup was used to measure the reflections from the MUTs. Sparse signal processing based signal decomposition with iterative dictionary updates was used to estimate the amplitudes and the phases of the reflected signal from the MUT. Based on the decomposition, the dielectric constant and the thickness of the MUT were estimated. The results 
show good agreement with the dielectric constants of the tested materials which are reported in the literature. Also, the thickness estimation error of the proposed approach $DU$ was always less than $2.95\%$ for the cases considered. Further, the iterative dictionary update approach improves thickness estimation compared to a fixed dictionary-based approach. In comparison, the $l^1-$norm minimization based methods are performing better compared to the $l^2-$norm minimization based method.



%

\section*{Acknowledgment}

The authors would like to thank Mr. Yamen Zantah and Mr. Benedikt Sievert for taking the VNA based measurements. The authors would like to thank Mr. Jochen Jebramcik and Mr. Orell Garten for their valuable discussions.

\ifCLASSOPTIONcaptionsoff
  \newpage
\fi

\bibliographystyle{IEEEtran}
\bibliography{ref}

%

\end{document}